\documentclass[a4paper,11pt]{article}
\pdfoutput=1 

\usepackage[utf8]{inputenc}
\usepackage{jinstpub}
\usepackage{pgf}
\usepackage[normalem]{ulem}
\RequirePackage{lineno}
\usepackage{color}
\usepackage{upgreek}
\usepackage{ulem}
\usepackage{nicefrac}

\newcommand{\PolA}{\mbox{${\alpha}$}}
\newcommand{\SphPol}{\mbox{${\theta}$}}
\newcommand{\SphAzi}{\mbox{${\varphi}$}}

\newcommand{\lumil}{\mbox{${\cal R}(d)$}}

\newcommand{\lumig}{\mbox{${\cal G}$}}
\newcommand{\trans}{\mbox{${\cal T}_{\SphPol_\text{in}=0}$}}

\begin{document}

\title{VUV Transmission of PTFE for Xenon-based Particle Detectors}

\affiliation[a]{Institut f\"ur Kernphysik, Westf\"alische Wilhelms-Universit\"at M\"unster, 48149 M\"unster, Germany}
\affiliation[b]{Physikalisches Institut, Universit\"at Freiburg, 79104 Freiburg, Germany}

\author[a]{Lutz~Althueser,}\emailAdd{l.althueser@uni-muenster.de}
\author[b]{Sebastian~Lindemann,}
\author[a]{Michael~Murra,}
\author[b]{Marc~Schumann,}
\author[a]{Christian~Wittweg,}
\author[a]{Christian~Weinheimer}

\abstract{Liquid xenon (LXe) based detectors for rare event searches in particle and astroparticle physics are optimized for high xenon scintillation light collection and low background rate from detector materials. Polytetrafluoroethylene (PTFE, Teflon\texttrademark) is commonly used to encapsulate the active LXe volume due to its high reflectance for VUV LXe scintillation light with peak emission at 178\,nm. Reflectance, transmission and number of background signals arising from PTFE depend on the thickness of the PTFE detector walls. In this work, we present VUV transmission measurements for PTFE of various thicknesses often considered in the design phase of LXe detectors. PTFE samples are measured in an apparatus previously used for reflectance measurements in LXe using collimated light at a wavelength of 178\,nm. Measurements in vacuum as well as gaseous xenon are described by the Kubelka and Munk model, as well as by Beer-Lambert's law for samples of $\geq 0.7\,\mathrm{mm}$ thickness, yielding a transmission coefficient of $\lambda_\text{BL} = (0.89 \pm 0.05)\,\mathrm{mm}$. The PTFE wall thickness of the XENONnT dark matter experiment was optimized by these measurements and selected as $\geq 3\,\mathrm{mm}$.}



\maketitle


\section{Introduction}
\label{sec::introduction}

Polytetrafluoroethylene (PTFE, Teflon\texttrademark) is commonly used in liquid xenon (LXe) based particle detectors owed to its excellent reflectivity for LXe scintillation light~\cite{yamashita, choi, bokeloh, levy, kaminsky, neves, silva, kravitz}, featuring peak emission in the vacuum ultraviolet (VUV) at 178\,nm~\cite{jortner}. LXe detectors are employed for rare event searches in particle and astroparticle physics, such as lepton flavor violation~\cite{meg}, neutrinoless double beta decay~\cite{exo, pandaxIII}, as well as dark matter searches~\cite{xe1t,lux,pandax}. The latter use dual-phase (liquid-gas) time projection chambers (TPC)~\cite{tpc} that provide position sensitivity, multi-scatter rejection, and allow for the identification of the interaction type (electron or nuclear recoil)~\cite{discr}. LXe-based TPCs are currently leading the search for dark matter in the form of weakly interacting massive particles (WIMPs) for almost all masses above 110\,MeV/c$^2$~\cite{xe1t,xe1tlowmass,Aprile:2019jmx}.

WIMPs are expected to produce a single scatter nuclear recoil signal~\cite{goodmanwitten}, thus neutron-induced nuclear recoils are one of the most critical backgrounds~\cite{review}. PTFE is a fluoropolymer (C$_2$F$_4$)$_n$ and contains two fluorine atoms per carbon atom. Fluorine only has one stable isotope, $^{19}$F, which has a high neutron emission yield in $(\upalpha,\,$n$)$ reactions~\cite{Fyield}. In an experimental setting $\upalpha$-particles originate from the primordial $^{238}$U and $^{232}$Th decay chains present in trace amounts in any material~\cite{heusser}. In order to maximize sensitivity it is crucial to minimize the amount of background signals arising from PTFE by minimizing its amount in a given detector. On the other hand, PTFE serves important purposes in the detectors: apart from maximizing the light collection efficiency~\cite{yamashita, xent} in the TPC, it is also used as insulating structural material~\cite{xe100instr, pandaxinstr, xe1tinstr} and optically separates the active LXe TPC (in which light and charge signals are recorded) from the inactive LXe surrounding the TPC. Scintillation light produced in this outer volume must not enter the TPC, as it would lead to artefacts such as accidental coincidences and could reduce the ability to distinguish electron and nuclear recoils. Accordingly, a balance between background reduction and optical shielding has to be found when determining the required thickness of the PTFE walls. This is especially important in the context of third generation LXe dark matter detectors such as DARWIN~\cite{darwinwimp,darwin}.

Here, we report on measurements of the transmission of collimated VUV light at a wavelength of 178\,nm through virgin grade PTFE in vacuum and gaseous xenon (GXe). Our results were used to optimize the thickness of the PTFE walls of the XENONnT TPC currently under commissioning~\cite{xent, ntmc}. Previous measurements indicated that the opacity increases with decreasing wavelength. However, no data was available below 400\,nm~\cite{tsai}.

The article is structured as follows: the experimental apparatus for the measurement is described in Chapter~\ref{sec::setup}. The measurements are detailed in Chapter~\ref{sec::analysis} and the data analysis and  results are presented in Chapter~\ref{sec::result}. The results are discussed in Chapter~\ref{sec::discussion}.


\section{M\"unster reflectance and transmission chamber}
\label{sec::setup}

\begin{figure}[bp]
\centering
\includegraphics[width=.9\textwidth]{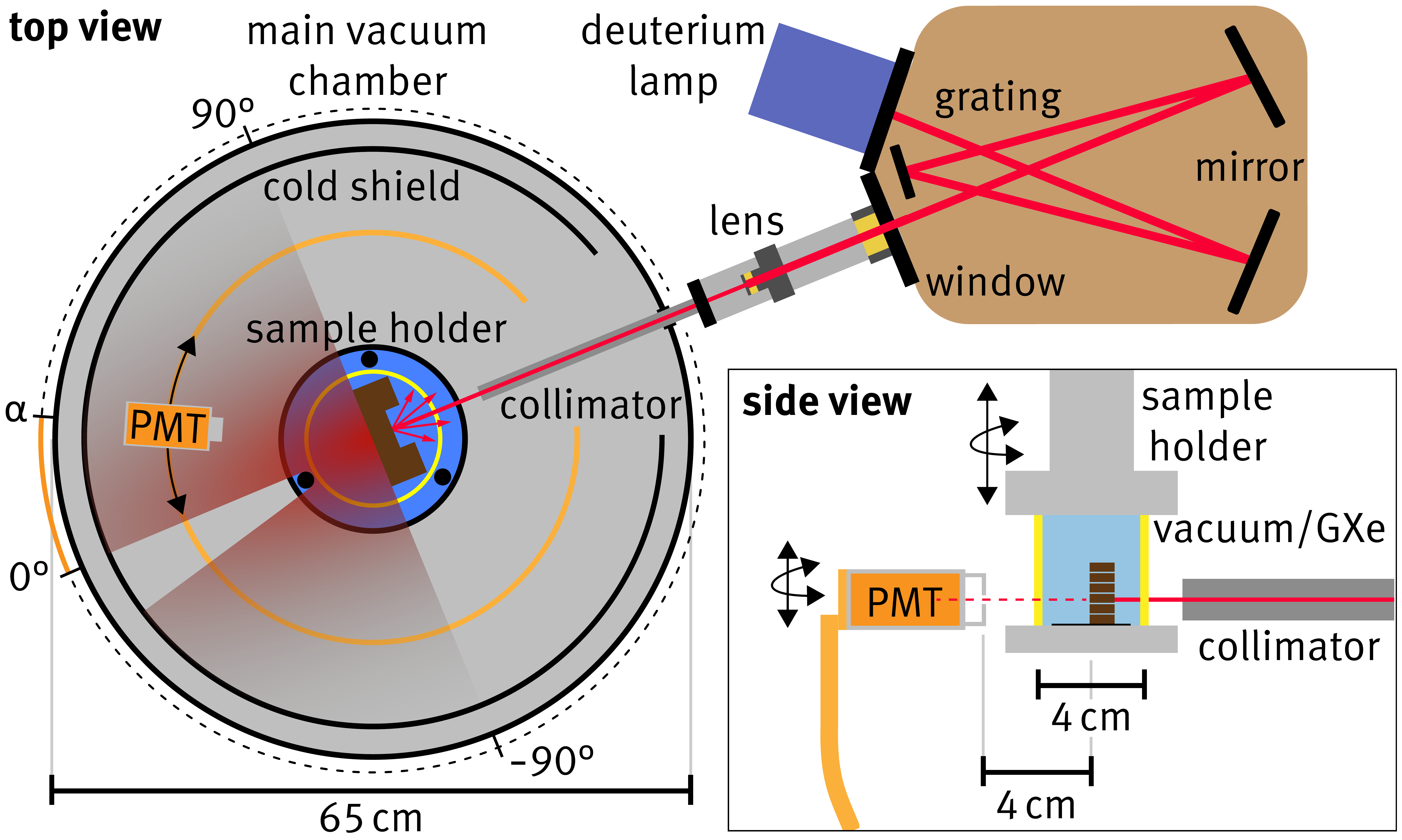}
\caption{\label{fig:setup} Schematic view of the experimental setup, focusing on components relevant for light transmission measurements. A narrow beam of VUV light around 178\,nm generated by a deuterium lamp and monochromator is guided onto a rotatable and height-adjustable PTFE sample. The PTFE sample is mounted inside of a quartz tube (yellow), that can be evacuated or filled with GXe. Transmitted light (red shaded) is measured with a PMT, equipped with an exchangeable aperture. The PMT can be moved along the PMT axis (orange) at various heights~$z$. Threaded rods, depicted as black dots in the top view, are blocking the light at certain azimuthal angles~$\PolA$.} 
\end{figure}

The apparatus to measure the transmission of VUV light around 178\,nm through PTFE is installed at the University of M\"unster and shown in Figure~\ref{fig:setup}. It was previously used for angle-resolved reflectance measurements off PTFE surfaces immersed in LXe~\cite{choi, bokeloh, levy, kaminsky, WAGENPFEIL2019577} and for photomultiplier tube (PMT) tests~\cite{PMTtests}.

VUV light is generated using a tunable deuterium lamp (McPherson Model~632) providing spectral lines in the range of 110\,nm to 170\,nm and continuous emission from 170\,nm to 400\,nm~\cite{McPDeuterium}. The wavelength of LXe scintillation light around 178\,nm is selected with a vacuum monochromator (McPherson Model~218) consisting of a series of mirrors and a 1200\,grooves/mm grating with a blaze wavelength of 150\,nm. Its resolution is specified as 5.3\,nm for a slit width of 2\,mm and a reciprocal linear dispersion for the grating of 26.5\,\AA/mm~\cite{bokeloh}: the light is continuously diffracted by the grating in the direction of the slit opening and a window of 5.3\,nm around 178\,nm is extracted from the continuous lamp spectrum. The diffraction causes a weak polarization of the initially unpolarized light~\cite{polarization}, however, at 178\,nm this effect is negligible for the measurements presented here. Light at 178\,nm has a mean free path of 3.7\,cm in air at normal pressure~\cite{bokeloh}, so the light source is evacuated using a turbo molecular pump (Pfeiffer TMU~071~P). The two main volumes of the instrument, VUV light source and main vacuum chamber, are decoupled by a UV-transmitting MgF$_2$~window with a transmission of 80$\%$ at 178\,nm~\cite{window}. The light is focused by a UV~lens and guided into the main vacuum chamber, accommodating the PTFE sample and light sensor, by a collimator with a circular aperture of 1\,mm diameter. The focus point of the light is in the center of the sample holder.

The main vacuum chamber, an O-ring sealed stainless steel cylinder of 65\,cm diameter, is equipped with a cooled 1''-square PMT sensitive to LXe scintillation light (Hamamatsu~R8520-406 PMT), the sample holder and a two-stage cold shield to reduce radiative heat influx. Scattering of light on the inner cold shield is reduced by using anodized and dyed black stainless steel. In order to allow measurements with LXe in the setup, PMT and cold shield are cooled by a cold head (Leybold RPK~1500~E), reaching 160\,K and 80\,K, respectively~\cite{bokeloh}. The sample holder is cooled by an Iwatani CryoMini (PD08) and kept at 193.15\,K, further reducing heat influx to the system. For the measurements presented here, cooling with both cold heads is used to reduce the PMT noise. Light deflection and condensation of water on cold surfaces, such as the PMT or sample holder, is prevented by evacuation down to $10^{-6}$\,mbar~\cite{levy} using a turbo molecular pump (Leybold TW~300). The pressure in both main volumes, and the temperatures of the laboratory, PMT, sample holder and cold shield are monitored during preparation and measurement phases using Pt100 sensors.

Light from the collimator is focused onto the PTFE sample and measured by the PMT. The PMT is mounted onto a copper support structure which can be rotated around the sample holder's center. The angular movement of the PMT has been automated using a step motor controlled by the data acquisition system. It is equipped with a 24-teeth gear wheel. A 96-teeth gear wheel is mounted on the PMT structure rotation feedthrough and connected by a rubber belt to the motor wheel. Thus, one step of the motor corresponds to 0.45° movement on the PMT rotation axis as illustrated in Figure~\ref{fig:setup}. Polar coordinates are used to describe the PMT movement in the main vacuum chamber: $\PolA$ is the polar or scattering angle of the PMT relative to the collimated light beam and $z$ is the height. PMT signals are amplified by a factor of 10 via a fast amplifier (CAEN~N979) and counted using a combination of a Leading Edge Discriminator (CAEN~N840) operated at a threshold of 70\,mV and a custom digital counter unit. The PMT signals have a typical width of 20\,ns and are converted to rectangular pulses of 40\,ns length by the discriminator. Additional signals within these 40\,ns are discarded and not counted. The relative dead time of $\mathcal{O}(30\,\text{kcps}) \cdot 40\,\text{ns} \approx 10^{-3}$ is negligible. The adjustment of the PMT operation voltage and temperature dependencies are discussed in~\cite{FSprenger, bokeloh}. The 20.5\,$\times$\,20.5\,mm$^2$ active area of the PMT is reduced to a circle of 1.5\,mm diameter using an aperture attached to the PMT. The distance of the PMT aperture opening and sample holder center is 4\,cm. Therefore, the solid angle covered by the PMT with respect to the PTFE sample amounts to $\Delta \Omega_\text{PMT} = 1.1$\,msr. Along the horizontal PMT rotation axis, the aperture covers $\Delta \PolA = 2.1$°.

The sample holder, placed in the center of the main vacuum chamber, can be rotated by 360° and adjusted in height by a motion feedthrough while the system is in operation~\cite{levy}. PTFE samples of various thicknesses are stacked, with a vertical distance of 7\,mm, and mounted on the holder allowing for consecutive measurements with comparable pressure, temperature and cleanliness conditions by adjusting the $z$-position of the sample holder, moving a specific sample into the VUV light beam. Samples can be measured either in vacuum or GXe --- as presented in this work --- or LXe. Measurements with xenon require to encapsulate the samples with a quartz tube as described in~\cite{levy, bokeloh}. The tube (from Quarzglas Heinrich) with a radius of 20\,mm and a wall thickness of 5\,mm is made of UV-grade fused silica and can withstand several bar over-pressure. It has a nominal light transmission of \textasciitilde80\% through the full tube at 178\,nm, crossing four surfaces. The refractive index of the tube at VUV wavelength is $\text{n} = 1.6$, thus \textasciitilde5\% of the light is reflected at each surface for photons with perpendicular incidence angle~\cite{levy}. The quartz tube is mounted using threaded rods, blocking transmitted and emitted light for certain PMT angles, as depicted in Figure~\ref{fig:setup}.

Samples of molded virgin grade high-density PTFE were machined from the same raw material. Each sample is 7\,mm high, 25\,mm wide and up to 5\,mm thick at the impact point of the collimated 178\,nm light beam.  The full width at half maximum (FWHM) of the light beam spot at the PTFE sample position is estimated to be about 1\,mm~\cite{bokeloh}. A circular mill-machined recess of 6\,mm diameter was machined around the impact point to leave PTFE walls of ``0\,mm'' (hole), 0.80\,mm, 1.00\,mm, 1.40\,mm, 2.15\,mm and 3.55\,mm thickness. The procedure to measure the depth of the recess, and thus the thickness~$d$ of the samples, is not sensitive to small variations over the recess area. Therefore, the recess could only be measured over the whole recess diameter, resulting in a systematic uncertainty of 0.05\,mm. The PTFE surfaces were not treated and thus not optimized for reflectivity. All samples are stacked on top of each other and fixed by means of screws. Light will be transmitted and emitted from the flat PTFE surface facing the PMT, allowing for measurements of the transmission profile at a selected PMT height.

\section{PTFE transmission measurements}
\label{sec::analysis}

In preparation for the actual transmission measurements, the alignment of PMT, sample holder and collimator is ensured with an optically visible light source from the inside of the monochromator by illuminating the PMT aperture through the ``0\,mm''-sample (hole). Nevertheless, a precise alignment of the sample holder with 178\,nm light is still necessary. Both, main vacuum chamber and monochromator volume, are evacuated to sufficiently low pressures of below $3 \times 10^{-8}$\,mbar and $7 \times 10^{-5}$\,mbar, respectively, before the deuterium lamp is turned on. The IONIVAC pressure gauge in the main vacuum chamber is turned off during measurements as it generates light during operational mode at low pressures~\cite{levy}. The PMT and cold shields are being cooled down while the deuterium lamp is reaching stable light emission at 178\,nm. The PMT is moved out of the beam center to a position at 25°, see Figure~\ref{fig:setup}. The PMT count rate during cooldown is monitored, revealing a combination of background and dark count rate of \textasciitilde100\,cps at room temperature and below \textasciitilde4\,cps at operation temperature with the selected data acquisition settings. The PTFE samples are passively cooled by the sample holder cold head which is kept at a constant temperature. Thus a constant temperature of the PTFE samples is assumed.

\begin{figure}[bp]
\centering
\includegraphics[width=.9\textwidth]{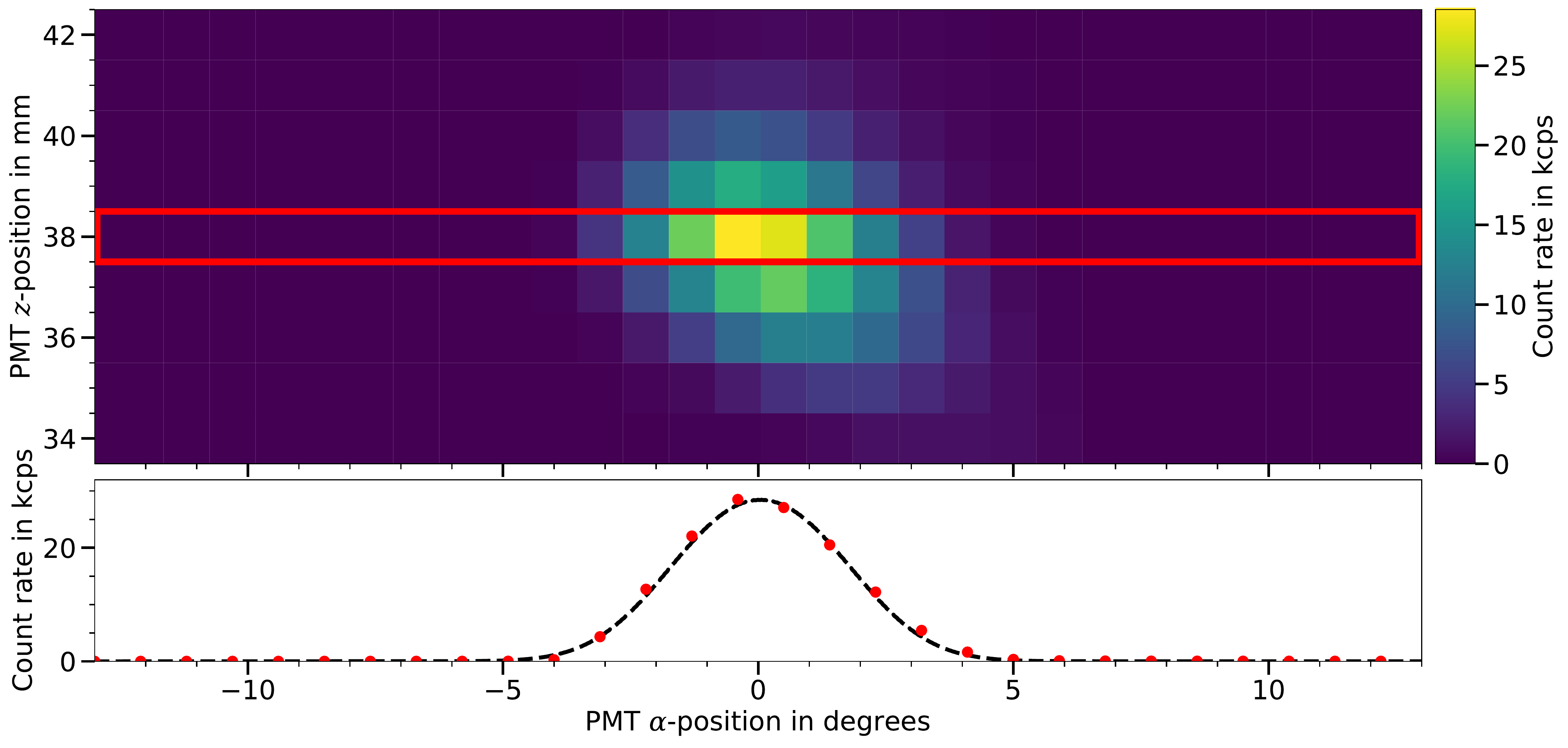}
\caption{\label{fig:beam} Two-dimensional profile of the collimated 178\,nm light beam measured in steps of $\Delta \PolA$\,=\,0.9° and $\Delta z$\,=\,1\,mm through the quartz tube and the ``0\,mm''-PTFE sample (hole) (upper panel) and horizontal scan through the two-dimensional profile at $z = 38\text{\,mm}$ (lower panel). The quartz tube is rotated such that the PTFE surface normal is pointing to $\PolA=-10$° and light is not blocked by the threaded rods. The horizontal scan is fitted by the Gaussian given in Equation~(\ref{eq:gaussian}).}
\end{figure}

The alignment of the collimated light beam, PTFE sample and PMT aperture in both, relative azimuthal angle~$\PolA$ and height~$z$, is a critical part of the measurement preparations. These alignment procedures are repeated for each measurement campaign. The sample holder is lifted out of the beam path and the PMT is aligned with the center of the collimated beam by measuring the light signal along the PMT axis for various PMT heights in $z$. Measurements in 0.9° steps from $\PolA=-20$° to $+$20° for each 1\,mm step in PMT height were taken. The center of the collimated 178\,nm light beam was obtained with an uncertainty of $\Delta \PolA=0.45$° and $\Delta z=0.5$\,mm. It exhibited a maximum count rate of \textasciitilde40\,kcps at a distance of 4\,cm to the focus point, in agreement with~\cite{levy}. Next, the ``0\,mm''-PTFE sample (hole) is moved into the beam path and the sample holder is rotated such that the PTFE surface normal is pointing to $\PolA=-10$°. Therefore, light is not blocked by the threaded rods. The light beam is measured again through the quartz tube surrounding the PTFE samples, leading to a similar two-dimensional distribution as without quartz tube. The profile is shown in Figure~\ref{fig:beam} and has a maximum count rate of \textasciitilde31\,kcps. Thus, the maximum count rate is decreased by the quartz tube to \textasciitilde78\% as expected for a light beam passing 4~times through a quartz-vacuum transition. As shown in the bottom plot of Figure~\ref{fig:beam}, a central one-dimensional slice through the two-dimensional intensity profile can be described by a Gaussian distribution $G_{38\,\text{mm}}(\PolA)$ with standard deviation $\sigma = 1.5$° and an amplitude of $G_0 = 31\,\text{kcps}$:
\begin{equation} 
   G_{38\,\text{mm}}(\PolA) = G_0 \cdot \exp{\left(- \frac{(\PolA - \PolA_0)^2}{2 \sigma^2}\right) }
   \text{.} \label{eq:gaussian}
\end{equation}
The standard deviation $\sigma$ describes the effective broadening of the light beam measured at the position of the PMT given by the angular divergence and the diameter of the light beam at the focal point as well as the diameter of the aperture in front of the PMT. 

The light beam can be directed through a PTFE sample of defined thickness by exact vertical ($z$-axis) positioning of the sample holder. The distance between two consecutive PTFE sample centers is fixed to 7\,mm by design. The precision for this positioning is estimated as 0.5\,mm. The effective thickness of the PTFE sample experienced by the collimated beam depends on the angle of the PTFE surface with respect to the beam. The angle of the sample holder is set to $90$° with respect to the collimator tube by means of a calibrated angle scale such that the PTFE surface normal points at $\PolA=0$°. This alignment was verified by measuring the collimated light through the ``0\,mm''-PTFE sample (hole) while rotating the sample holder. Measurements in vacuum and GXe were taken following the exact same procedure.

\begin{figure}[tbp]
\centering
\includegraphics[width=.9\textwidth]{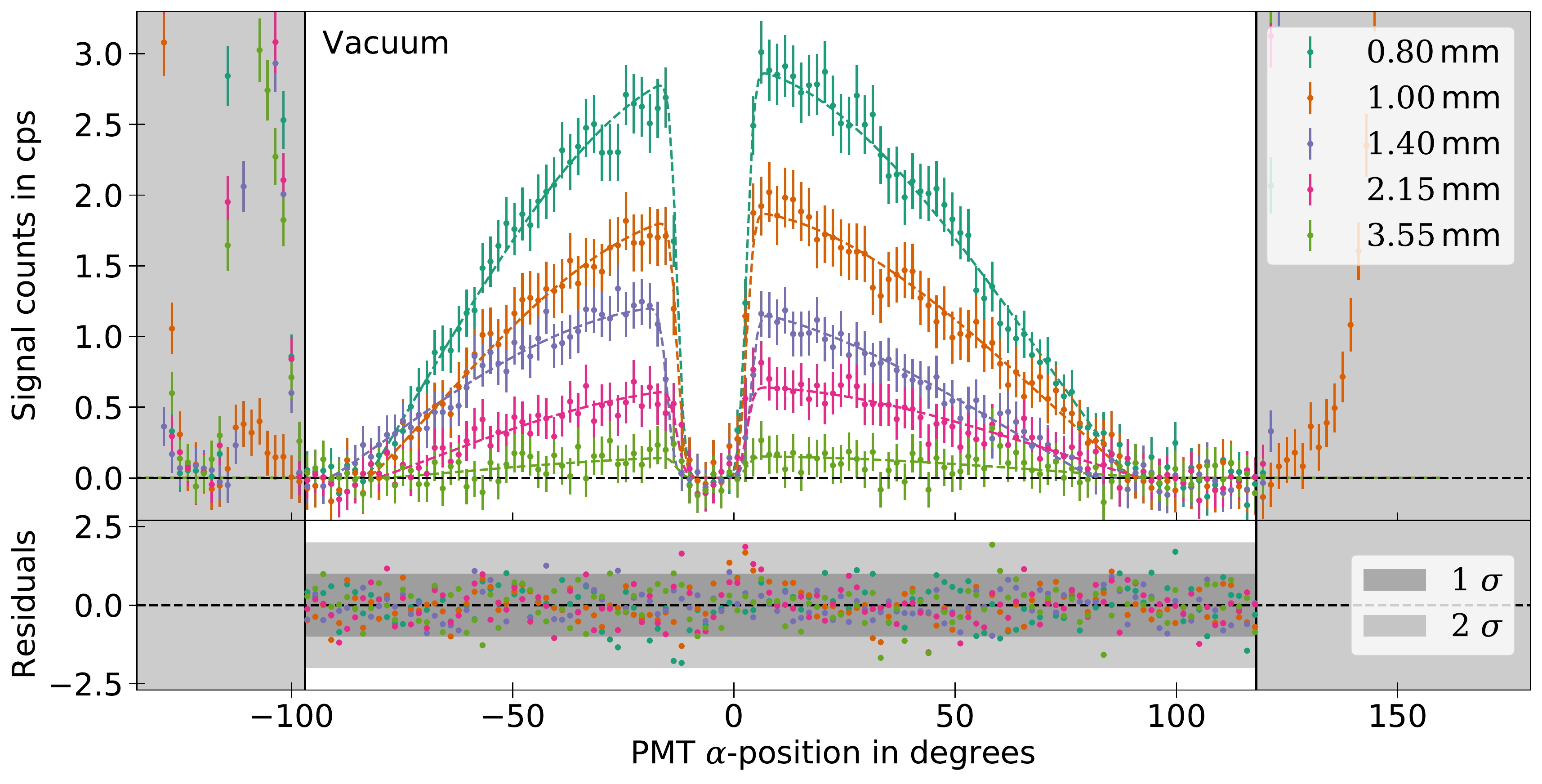}
\includegraphics[width=.9\textwidth]{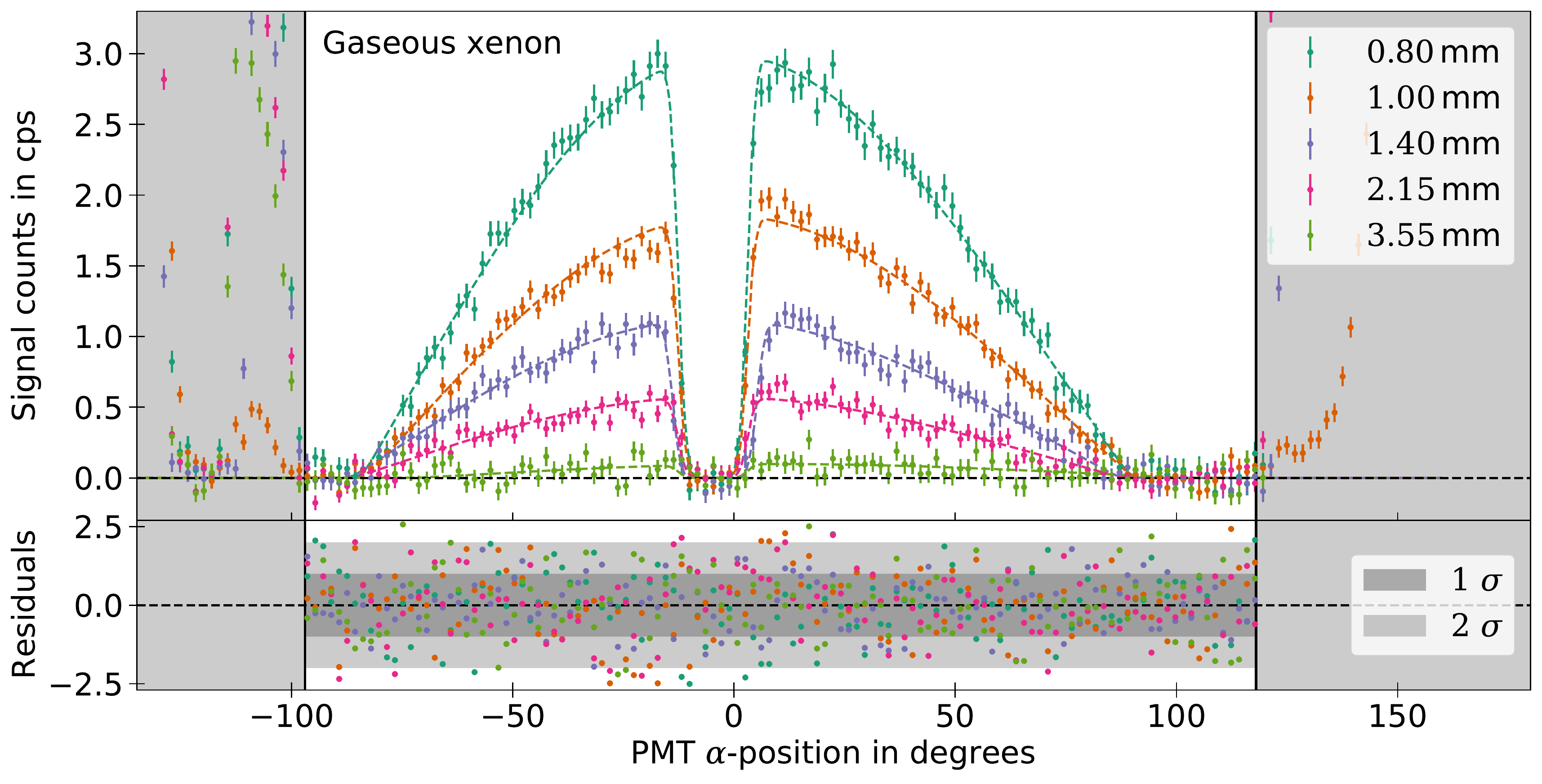}
\caption{\label{fig:measurement} Measurements of various PTFE samples (thickness indicated in different colors) encapsulated in a quartz tube surrounded by vacuum (upper panel) and 1\,bar gaseous xenon (lower panel). Statistical uncertainties are shown for each data point. The signal count rates (dashed lines) are obtained by subtracting the fitted background rates $B(\PolA, a, b)$ from the total count rates $I(\PolA, d)$. Shaded areas depict PMT angles at which photons reflected from the PTFE surface are expected to be detected. At $-$125°, $-$5° and 115° photons are blocked and partly reflected by threaded rods such that they are not detected. The PTFE samples are kept at $-$80\,°C for all measurements.}
\end{figure}

Several scans over the full range of PMT angular positions are taken for each PTFE sample, depicted as PMT axis in Figure~\ref{fig:setup}. The PMT height is kept at the $z$-position of the beam maximum. Each scan covers a range from $-$145° to $+$170°, see Figure~\ref{fig:measurement}.  Data points are taken every 1.8° corresponding to 4~steps of the stepper motor. Each PMT position is measured for 120\,s, while the counts of the first 4\,s are excluded in the analysis to avoid a possible impact of small mechanical vibrations after the PMT movement. The PMT window and aperture opening face the PTFE sample center at any angle. Therefore, photons reach the photocathode at nearly 0°~incident angle at the center of the PMT. Differences of the PMT output in dependence of the photon incident angle and position on the photocathode can thus be neglected. The absolute rate of photons detected is reduced compared to the number of photons transmitted for all measurements due to the PMT detection efficiency, defined by the combination of PMT quantum efficiency and collection efficiency.

Up to 10~complete scans per PTFE sample are combined by calculating the total number of counts and measurement time to obtain the rate of transmitted photons detected by the PMT. As shown in Figure~\ref{fig:measurement}, for angles below $-$95° and above 115° photons are reflected off the PTFE surface and absorbed in the main chamber. No transmitted photons are detected at the PTFE sample sides at $\pm$90°, as well as at the angular positions of the threaded rods (around $-$125°, $-$5° and 115°). The total count rate $I(\PolA, d)$ at an angle $\PolA$ is given by:
\begin{align}
    I(\PolA, d)
    &= (f_\text{shadow}(\PolA) \cdot R(\PolA, d)) \otimes G_0'(\PolA) + B(\PolA, a, b)  \nonumber \\
    &= 
    \int_0^{\pi/2} f_\text{shadow}(\PolA') \cdot I_0(d) \cdot \cos(\PolA') \cdot \frac{1}{\sqrt{2\pi} \sigma'}\exp\left(-\frac{(\PolA-\PolA')^2}{2 \sigma'^2}\right) d\PolA' 
    + (a \cdot \PolA + b) \text{.} \label{eq:fit_function}
\end{align}
$R(\PolA, d) = I_0(d) \cdot \cos(\PolA)$ is the rate of transmitted photons and $B(\PolA, a, b) = a \cdot \PolA + b$ the background count rate. The rate of transmitted photons $R(\PolA, d)$ is described by Lambert's cosine law with an intensity in the direction of the surface normal $I_0(d)$ for a PTFE sample of thickness~$d$. $R(\PolA, d)$ is multiplied by the additional function $f_\text{shadow}(\PolA)$ to account for the blocking of the light beam by the threaded rods, of which the central one is blocking photons in the angular range of $-$12.6° to 3.0°: $f_\text{shadow}(\PolA)$ is zero in the angular  direction of the threaded rods and unity otherwise. Because of the diameter of the light beam at the PTFE sample and the aperture in front of the PMT, the shadow of the threaded rods is not infinitely sharp. Thus $R(\PolA, d)$ is convoluted  with the normalized Gaussian~$G_0'(\PolA)$. Its standard deviation~$\sigma' = 1.4$° differs from~$\sigma$ in Equation~(\ref{eq:gaussian}), as --- in contrast to the three contributions to the standard deviation~$\sigma$ of the measured beam profile $G_{38\,\text{mm}}(\PolA)$ --- only the diameter of the light beam at the PTFE sample and the aperture in front of the PMT contribute to the broadening of the shadow of the threaded rods but not the angular divergence of the light beam. These relations between $G_{38\,\text{mm}}(\PolA)$ and $G_0'(\PolA)$ were confirmed by a toy Monte Carlo simulation.
The background rate $B(\PolA, a, b)$ consists of a temperature-dependent dark count rate and a constant stray light contribution. The dark count rate was measured to change linearly with time due to a slowly changing laboratory temperature. As the PMT angle $\PolA$ is selected sequentially for each measurement of 120\,s duration, this leads to a dark count rate model which depends linearly on $\PolA$. 

Not shown in Equation~(\ref{eq:fit_function}) is a correction of the angle $\PolA$ for misalignments of the PTFE samples in the sample holder with respect to the PMT aperture. These misalignments in both directions within the scattering plane (\emph{top view} in Figure~\ref{fig:setup}) are about 2\,mm in size and equal for all scans. They account for both, a shift of the PTFE sample surface with respect to the center of the sample holder and an offset of the PMT rotation around the sample holder center. 

\begin{figure}[tp]
\centering
\includegraphics[width=.9\textwidth]{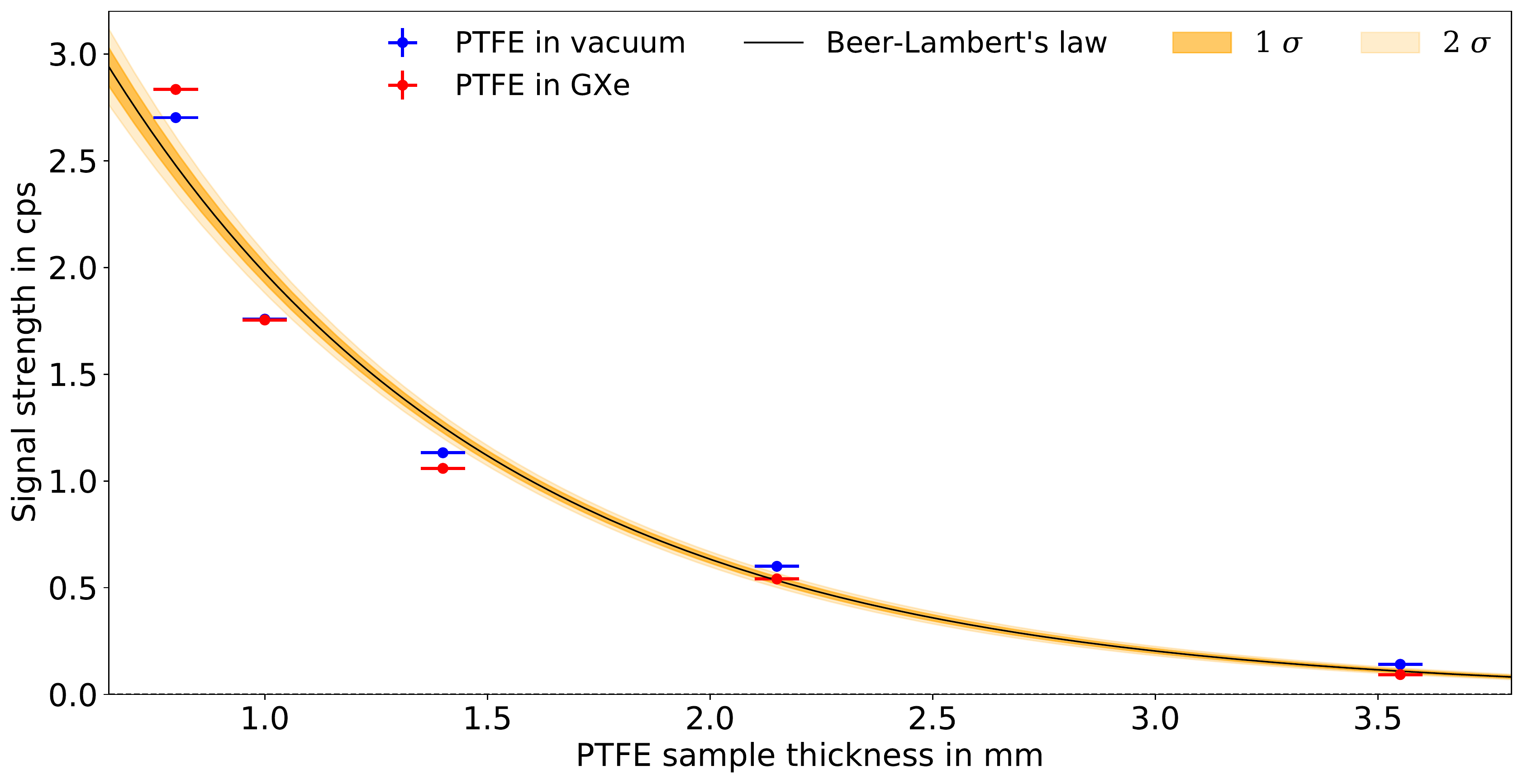}
\caption{\label{fig:result} Signal amplitudes $I_0(d)$ of the transmission measurements $R(\PolA, d)$ for various PTFE samples, fitted with Beer-Lambert's law. Statistical uncertainties are shown for the signal amplitudes and systematical measurement uncertainties for the PTFE sample thickness. The transmission parameters are estimated to be $I_0(0) = (6.1 \pm 0.6)$\,cps and $\lambda_\text{BL} = (0.89 \pm 0.05)$\,mm.}
\end{figure}

The measured count rates for all measurements shown in Figure~\ref{fig:measurement} are simultaneously fitted by $\chi^2$-minimization using $I(\PolA, d)$, sharing the parameters for $G_0'(\PolA)$, the angles $\PolA$ of the threaded rods and the misalignment correction. The background rate parameters~$a$ and~$b$ are determined by the simultaneous fit of all measurements for each sample, and range from 0.2\,mcps/° to 0.7\,mcps/° and 1.8\,cps to 4.0\,cps, respectively. Therefore, the maximum difference of the first and last angular position in the background rate $B(\PolA, a, b)$ of a single scan is~0.14\,cps. Signal amplitudes~$I_0(d)$ are shown for each measurement in Figure~\ref{fig:result}.

The alignment of the PMT axis with the collimated light beam in $\PolA$ and $z$ was checked before and after each measurement campaign. No shift in the PMT positioning and collimator direction could be observed. However, the alignment of the PTFE surface normal could only be verified in $\PolA$ direction. A tilt $\Theta$ of the PTFE samples stacked on the sample holder would lead to a systematic shift of the full hemisphere of transmitted light. The observed photon rates would be reduced by a factor of $1-\cos(\Theta)$, which corresponds to a negligible reduction of 0.4\% for a conservative estimate of $\Theta=5$°. The potential systematical effect due to a changing time-variable light intensity was mitigated by measuring PTFE samples of various thickness in random order and by measuring the beam profile $G_{38\,\text{mm}}(\PolA)$ before and after the measurements in vacuum and GXe, which were found in agreement.

\section{Results}
\label{sec::result}
The transmission of VUV light through PTFE can be described by Beer-Lambert's law or the Kubelka and Munk model~\cite{Kubelka1931, Kubelka1948, Kubelka1954, Boroumand}, making certain assumptions for each. Incident light can be either reflected on the PTFE surface or scattered in the bulk material and then reflected, absorbed or transmitted. Light reflected at the surface include a broadened component around the direction of the specular reflection due to the surface roughness. The remaining incident light penetrates the sample and scatters in the bulk material. The re-emission of light is described as diffuse reflection and is not dependent on the incident angle. Scattered light can be reflected, but also be absorbed by or transmitted through the PTFE. Therefore, for thin samples the absolute reflectivity is expected to increase with the PTFE thickness. Consequently, light transmission decreases with increasing (thin) sample thickness depending on the depth at which the scattering processes dominate.

Beer-Lambert's law is applicable if light is only reflected at the PTFE surface and absorbed within the bulk material described by the transmission coefficient~$\lambda_\text{BL}$. Additional scattering processes in the bulk material are not taken into account. Reflectivity measurements of the PTFE samples were not the scope of this work but measurements of comparable PTFE samples show no evidence for changes of the reflectivity above 1\,mm thickness~\cite{Haefner, levy}. Hence, Beer-Lambert's law is assumed to be applicable for modeling the amount of light~$I_0(d)$ transmitted through PTFE samples of thickness~$\gtrsim$1.0\,mm given by:
\begin{equation}
    I_{\text{BL}}(d) = I_0(d) = I_0(0) \cdot \exp{\left( - \frac{d}{\lambda_\text{BL}} \right) } \text{,} \label{eq:beer}
\end{equation}
where $I_0(0)$ represents the signal strength of light transmitted through an infinitesimally thin PTFE sample, which is expected to reflect and transmit but not absorb light. $\lambda_\text{BL}$ is the transmission coefficient of the medium. Both parameters are obtained for the PTFE samples in vacuum and GXe separately by $\chi^2$-minimization to $I_\text{0, vacuum}(0) = (5.7 \pm 0.7)$\,cps, 
$\lambda_\text{BL, vacuum} = (0.94 \pm 0.07)$\,mm,
$I_\text{0, GXe}(0) = (6.7 \pm 1.1)$\,cps 
and $\lambda_\text{BL, GXe} = (0.82 \pm 0.07)$\,mm. The parameters $\lambda_\text{BL, GXe}$ and $\lambda_\text{BL, vacuum}$ agree within their uncertainties, as expected due to the similar refractive indices for vacuum and GXe ($\text{n}_{178\text{\,nm}} = 1.00109$)~\cite{xegasn}. Thus, measurements in vacuum and GXe are combined, yielding
\begin{equation}
  I_0(0) = (6.1 \pm 0.6)\,\text{cps \qquad and \qquad } \lambda_\text{BL} = (0.89 \pm 0.05)\,\text{mm.} \nonumber
\end{equation}
The absolute light transmission for various thicknesses can be obtained by integrating the light beam profile $G_{z}(\PolA)$ and rate of transmitted photons $R(\PolA, d)$ over the full hemisphere behind the PTFE sample. Both functions are converted into spherical coordinates with the polar angle~$\SphPol$ and the azimuthal angle~$\SphAzi$. The light transmission is assumed to be symmetric around the collimated light beam direction for scattering angles $\SphPol$. The solid angle of the detector $\Delta \Omega_\text{PMT}$ under which all rates have been measured needs to be taken into account to obtain the correct total rates. The total radiances of the collimated light beam \lumig~hitting the PTFE samples and of the transmitted photons \lumil~are both measured through the quartz tube under identical conditions and given by:
\begin{align}
    \lumig &= \frac{1}{\Delta \Omega_\text{PMT}} \cdot \int_{0}^{2\pi} \int_{0}^{\pi/2} G_{\SphAzi}(\SphPol) \cdot \sin(\SphPol) \,d\SphPol d\SphAzi \textnormal{,}\\
    \lumil &= \frac{1}{\Delta \Omega_\text{PMT}} \cdot \int_{0}^{2\pi} \int_{0}^{\pi/2} R(\SphPol, d) \cdot \sin(\SphPol) \,d\SphPol d\SphAzi \nonumber\\
    &= \frac{2\pi \cdot I_0(d)}{\Delta \Omega_\text{PMT}} \cdot \int_{0}^{\pi/2} \cos(\SphPol) \sin(\SphPol) \,d\SphPol = 
    \frac{\pi \cdot I_0(d)}{\Delta \Omega_\text{PMT}}
    \text{.}
\end{align}
The total transmission probability $\trans (d)$ for collimated light at perpendicular incidence can be defined as
\begin{align}
     \trans (d) = \frac{{\cal R}(d)}{{\cal G}} \text{,}
     \label{eq:lumi}
\end{align}
yielding transmission probabilities of $\trans (0.8\text{\,mm})= (0.042 \pm 0.002)$, $\trans (3.0\text{\,mm})= (0.0034 \pm 0.0002)$ and $\trans (5.0\text{\,mm})= (0.0004 \pm 0.0001)$ for the PTFE samples used in this work. The total radiance of the collimated light beam $\lumig = (185 \pm 9)\text{~cps}/\Delta \Omega_\text{PMT}$ and the transmitted radiance extrapolated for vanishing PTFE thickness $\lim\limits_{d \rightarrow 0} \lumil = (19.3 \pm 0.9)\text{~cps}/\Delta \Omega_\text{PMT}$ differ by about a factor 10 using the Beer-Lambert's law. 

\begin{figure}[bp]
\centering
\includegraphics[width=.9\textwidth]{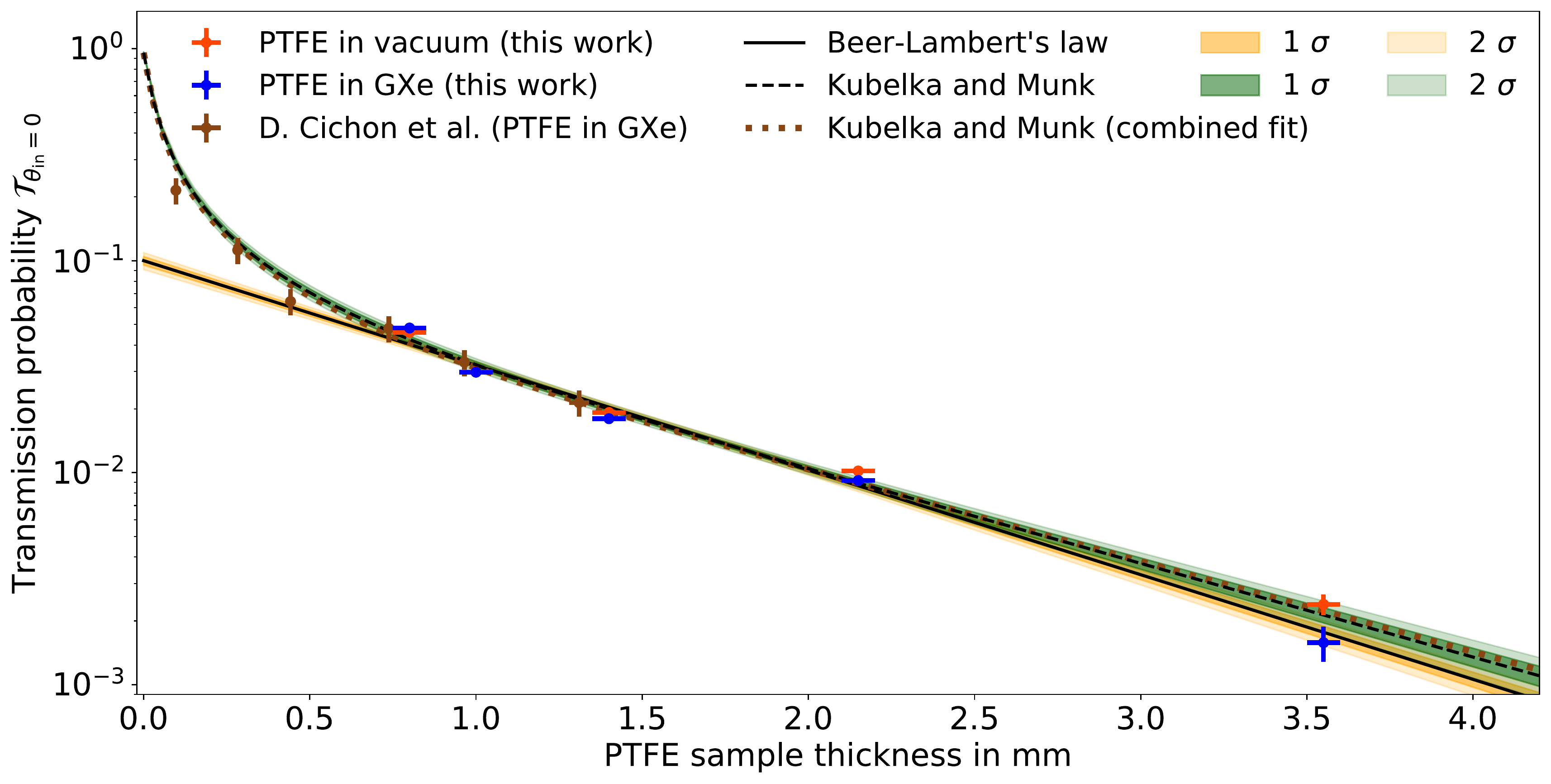}
\caption{\label{fig:result_KM} Transmission probabilities $\trans (d)$ of the signal amplitudes $I_0(d)$ for various PTFE samples and the converted Beer-Lambert's law (solid line), as given in Figure~\ref{fig:result}. Measurements in vacuum and GXe are combined and fitted by the Kubelka and Munk model (dashed line), yielding $\lambda_K = (46 \pm 6)\,$mm, $\lambda_S = (0.042 \pm 0.002)\,$mm and $\lambda_{\text{BL}_\text{KM}} = (0.98 \pm 0.05)\,$mm. Measurements of comparable PTFE samples reported by Cichon et al.~\cite{heidelberg} (PTFE in GXe, \emph{room temperature setup}) are combined with the vacuum/GXe measurements of this work (dotted line), yielding $\lambda_K = (52 \pm 6)\,$mm, $\lambda_S = (0.040 \pm 0.002)\,$mm and $\lambda_{\text{BL}_\text{KM}} = (1.02 \pm 0.05)\,$mm.}
\end{figure}

The \textit{two-flux} model originally proposed by Kubelka and Munk~\cite{Kubelka1931, Kubelka1948, Kubelka1954, Boroumand} includes the diffuse light scattering processes in the bulk material and can be used to resolve this discrepancy for vanishing PTFE thickness $d\to 0$. Diffuse reflectance is modeled by a forward- and a backward-scattering component for each sample layer. Specular reflections at the sample surface are not included in the original Kubelka and Munk formula. The diffuse transmittance for a finite sample thickness $d$ is given by
\begin{align}
     T_{\SphPol_\text{in}=0} (d) &= \frac{\mathcal{T}_{\SphPol_\text{in}=0} (d)}{1 - \text{R}_F} = \frac{b}{a \cdot \sinh(b \cdot S \cdot d) + b \cdot \cosh(b \cdot S \cdot d)} \text{.}
     \label{eq:KMTrans}
\end{align}
The parameter $a = 1 + \nicefrac{K}{S}$ and $b = \sqrt{a^2 - 1}$ are defined by the effective absorption coefficient $K = \nicefrac{1}{\lambda_K}$ and the effective scattering coefficient $S = \nicefrac{1}{\lambda_S}$. This formulation of the Kubelka and Munk model assumes that no transmitted light is scattered back into the sample. Measured transmission probabilities~$\trans (d)$ have to be corrected by the fraction of light $\text{R}_F$ which is specularly reflected at the surface. The specular reflection component $\text{R}_F$ is unknown for the samples used in this work and can be calculated using the refractive index $\text{n}_\text{PTFE}$ following Fresnel's law for perpendicular incidence ($\SphPol_\text{in}=0$). The PTFE refractive index is chosen as $\text{n}_\text{PTFE} = (1.52 \pm 0.15)$ including a 10$\%$ systematic uncertainty to cover the range of literature values from $1.39$ to $1.66$ at 178\,nm~\cite{kravitz}. Thus the specular reflection component for perpendicular incidence is given by
\begin{equation}
    \text{R}_{F} = \left( \frac{\text{n}_\text{GXe, vacuum}-\text{n}_\text{PTFE}}{\text{n}_\text{GXe, vacuum}+\text{n}_\text{PTFE}} \right)^2 = (0.04 \pm 0.02) \text{.}
\end{equation}
Vacuum and GXe measurements are converted to measured transmission probabilities~$\trans (d)$ by using the total radiance of the collimated light beam~$\lumig$ and are given in Figure~\ref{fig:result_KM}. A $\chi^2$-minimization fit of $T_{\SphPol_\text{in}=0} (d)$ to the combined vacuum and GXe data yields
\begin{equation}
  \lambda_K = (46 \pm 6)\,\text{mm \qquad and \qquad } \lambda_S = (0.042 \pm 0.002)\,\text{mm.} \nonumber 
\end{equation}
The parameter $\text{R}_{F}$ was treated as a free fit parameter constrained by the given systematic uncertainty. Beer-Lambert's law and the Kubelka and Munk model can be directly compared by using the Kubelka and Munk approximation for intensely scattering materials with $(S \cdot d) \rightarrow \infty$ and $\text{R}_d < 1$ from~\cite{Kubelka1948}:
\begin{align}
    T_{\SphPol_\text{in}=0} (d) &\approx 2b \cdot \text{R}_d \cdot \exp{\left( -b \cdot S \cdot d \right)} \nonumber \\
    &= 2b \cdot \text{R}_d \cdot \exp{\left( -\frac{d}{\lambda_{\text{BL}_\text{KM}}} \right)} \text{,} \label{eq:KMTransapprox}\\
    \lambda_{\text{BL}_\text{KM}} &= \frac{1}{b \cdot S} \stackrel{\lambda_K \gg \lambda_S}{\approx} \sqrt{\frac{\lambda_K \cdot \lambda_S}{2}} \\
    &= (0.98 \pm 0.05)\,\text{mm,} \nonumber
\end{align}
where $\text{R}_d$ is the diffuse reflectance and $\lambda_{\text{BL}_\text{KM}}$ the effective transmission coefficient of the medium. Equation~(\ref{eq:KMTransapprox}) is only applicable for sample thicknesses $d$ greater than the depth at which
the scattering processes dominate. Beer-Lambert's law and the Kubelka and Munk model are in agreement for sample thicknesses $d \geq 0.7\,$mm. Thinner samples can only be described by the Kubelka and Munk model, indicating that the diffuse reflectance is reduced and that Equation~(\ref{eq:KMTransapprox}) is not applicable in that range. The transmission probability at perpendicular incidence~$T_{\SphPol_\text{in}=0} (d)$ for a sample thickness~$d$ is the upper limit for the light transmission for all incidence angles~$\SphPol_\text{in}$. The specular reflection component~$\text{R}_F$ dominates for high incidence angles~\cite{kravitz} and will reduce the overall light transmission, depending on the specific distribution of~$\SphPol_\text{in}$.


\section{Discussion}
\label{sec::discussion}
In this work we have shown that the experimental setup, previously used for reflectance measurements reported in~\cite{bokeloh, levy, kaminsky, WAGENPFEIL2019577}, can also be used to measure the transmission of materials at the wavelength of LXe scintillation light. These measurements were taken using collimated light around 178\,nm with 0°~incidence angle and a PMT orientation equal to the direction of the transmitted photon for all accessible angles. The light transmission of molded virgin grade high-density PTFE is described by two models for samples at about $-$80\,°C in vacuum and GXe. It was found that Beer-Lambert's law with a transmission coefficient $\lambda_\text{BL} = (0.89 \pm 0.05)$\,mm is applicable for PTFE samples of thickness $d \geq 0.7\,$mm while the Kubelka and Munk model with the effective coefficients~$\lambda_K = (46 \pm 6)\,$mm and~$\lambda_S = (0.042 \pm 0.002)\,$mm is valid for all thicknesses. An approximation of the Kubelka and Munk model yields $\lambda_{\text{BL}_\text{KM}} = (0.98 \pm 0.05)\,$mm which is in agreement with Beer-Lambert's law. The effective coefficients~$\lambda_K$ and~$\lambda_S$ are expected to depend on the surrounding medium and on the temperature of the PTFE samples due to thermal density and structural changes. It was found that the measurements for vacuum and GXe are in agreement, as expected from their similar refractive indices.
The measurements presented here benefit from well-known and quantifiable systematic uncertainties, the fact that the collimated VUV light passes only once through the PTFE samples, and the ability for angle-resolved measurements. By using a well-motivated model, which is supported by the measured data, the angular information allows us to extrapolate the transmitted light signal into regions which were experimentally not directly accessible in order to ensure that no transmitted light is left unaccounted for in the analysis.

Independent measurements of xenon scintillation light transmission through PTFE were performed recently~\cite{heidelberg} with the PTFE samples being immersed in GXe at room temperature as well as in LXe at cryogenic temperatures. After combining both data sets the authors obtain an about two times smaller transmission coefficient~$\lambda$ applying Beer-Lambert's law, which significantly disagrees with the one presented here. However, as seen in Figure~\ref{fig:result_KM}, Beer-Lambert's law is not applicable for sample thicknesses investigated in \cite{heidelberg}. Applying the Kubelka and Munk model, which also holds for thinner PTFE thicknesses, shows that the measurements reported in~\cite{heidelberg} and in this work are in agreement. A combined fit using the Kubelka and Munk model yields the coefficients $\lambda_K = (52 \pm 6)\,$mm, $\lambda_S = (0.040 \pm 0.002)\,$mm and $\lambda_{\text{BL}_\text{KM}} = (1.02 \pm 0.05)\,$mm.

PTFE components are used to encapsulate the inner active volume in LXe detectors from the outer non-active LXe volume, equipped with passive detector components like electric field shaping rings and cables. Scintillation light created outside the active volume, e.g., by $\upalpha$-decays, must be hindered from entering the inner active volume where it could contribute to accidental coincidence signals, or --- if $T_{\SphPol_\text{in}=0}$ would be too large --- lead to leakage of background events into the signal region. Light generated in the outer LXe region can be assumed as VUV scintillation photons isotropically emitted from a point-like source yielding a sinusoidal distribution of the angles of incidence. The total hemispherical reflectance of PTFE is constant for most incident angles and increases slightly for higher angles~\cite{kravitz}, however, the fraction of light reflected following Fresnel's law at the PTFE surface dominates for higher incident angles. As a consequence, less light will penetrate the PTFE and the transmission probability will decrease compared to the measurements for perpendicular incidence presented here. The effective path through an infinitely large PTFE wall for all incident angles will be dominated by scattering processes in the bulk material and follow the observed Beer-Lambert's law or Kubelka and Munk model such that $T_{\text{isotropic}} (d) < T_{\SphPol_\text{in}=0} (d)$ can be assumed.

Since PTFE walls of LXe detectors are usually polished to further enhance the absolute reflectance, unlike the PTFE samples used in this work, the absolute reflectance and especially the specular component can be assumed to be higher resulting in the product of total transmission ${\cal R}(d)/{\cal G}$ to be lower. In addition, PTFE immersed in LXe will further increase the absolute reflectance to well above 90$\%$~\cite{chepel, kravitz, Haefner}, while PTFE samples measured in vacuum or GXe -- as in this work -- showed an absolute reflectance in the range of 55-70\%~\cite{Haefner,levy}. These effects will reduce the absolute transmission $T_{\text{isotropic}} (3\,\text{mm})$ further into the low $10^{-4}$ range and the fraction of the primary $\upalpha$-decay light (typically $4 - 6\,\text{MeV}$) penetrating into the active detector volume will be below the detector threshold (typically $\sim$1\,keV). Most LXe detectors feature a complex detector design in the outer LXe volume severely reducing the light collection efficiency owed to absorption on other materials and PTFE wall components of partially increased thickness (e.g., for structural purposes). These effects of significantly reducing the amount of transmitted light are neglected in this conservative estimate. Also neglected is the fact that any accidental light signal transmitted through a PTFE wall of finite thickness needs to be paired with a suitable charge signal in dual-phase TPCs such that it can be detected and contribute to the background rate.

The XENONnT direct dark matter experiment~\cite{xent} was designed with the goal of having a PTFE wall as thin as possible for separating the inner active LXe volume from the outer non-active LXe volume in order to reduce the radioactivity originating from the PTFE wall itself while still optically separating the active from the inactive LXe volume. The measurements reported in this work have led to the decision to use PTFE walls of at least 3\,mm thickness compared to the 5\,mm used on previous XENON detectors~\cite{xe100instr,xe1tinstr}.

\section*{Acknowledgments}
This work is supported by DFG through GRK~2149: Strong and Weak Interactions --- from Hadrons to Dark Matter and by the European Research Council (ERC) grant No.~724320 (ULTIMATE).

\bibliography{main.bib}{}

\providecommand{\href}[2]{#2}\begingroup\raggedright\begin{thebibliography}{10}

\bibitem{yamashita}
M.~Yamashita, T.~Doke, K.~Kawasaki et~al., \emph{Scintillation response of
  liquid {Xe} surrounded by {PTFE} reflector for gamma rays},
  \href{https://doi.org/10.1016/s0168-9002(04)01663-8}{\emph{Nucl. Instrum.
  Methods Phys. Res., Sect. A} {\bfseries 535} (2004) 692}.

\bibitem{choi}
B.~Choi, \emph{The Light Response of the XENON100 Time Projection Chamber and
  the Measurements of the Optical Parameters with the Xenon Scintillation
  Light}, Ph.D. thesis, Columbia University, 2013.
\newblock
  \href{https://doi.org/10.7916/D84Q8262}{https://doi.org/10.7916/D84Q8262}.

\bibitem{bokeloh}
K.~Bokeloh, \emph{Calibration of hot and cold dark matter experiments: an
  angular-selective photoelectron source for the KATRIN experiment and an
  apparatus to determine the reflection properties of PTFE for vacuum UV
  light}, Ph.D. thesis, WWU M\"unster, 2013.
\newblock
  \href{https://nbn-resolving.org/urn:nbn:de:hbz:6-34329395924}{https://nbn-resolving.org/urn:nbn:de:hbz:6-34329395924}.

\bibitem{levy}
C.~Levy, \emph{Light Propagation and Reflection off Teflon in Liquid Xenon
  Detectors for the XENON100 and XENON1T Dark Matter Experiment}, Ph.D. thesis,
  WWU M\"unster, 2014.
\newblock
  \href{https://nbn-resolving.org/urn:nbn:de:hbz:6-32399368441}{https://nbn-resolving.org/urn:nbn:de:hbz:6-32399368441}.

\bibitem{kaminsky}
J.~B. Kaminsky, \emph{Optimizing liquid Xenon TPCs}, Ph.D. thesis,
  Universit{\"a}t Bern, 2017.
\newblock
  \href{https://doi.org/10.7892/boris.134683}{https://doi.org/10.7892/boris.134683}.

\bibitem{neves}
F.~Neves, A.~Lindote, A.~Morozov et~al., \emph{Measurement of the absolute
  reflectance of polytetrafluoroethylene ({PTFE}) immersed in liquid xenon},
  \href{https://doi.org/10.1088/1748-0221/12/01/p01017}{\emph{Journal of
  Instrumentation} {\bfseries 12} (2017) P01017}
  [\href{https://arxiv.org/abs/1612.07965v1}{{\ttfamily 1612.07965v1}}].

\bibitem{silva}
C.~Silva, J.~Pinto~da Cunha, A.~Pereira et~al., \emph{Reflectance of
  polytetrafluoroethylene for xenon scintillation light},
  \href{https://doi.org/10.1063/1.3318681}{\emph{Journal of Applied Physics}
  {\bfseries 107} (2010) 064902}
  [\href{https://arxiv.org/abs/0910.1056v1}{{\ttfamily 0910.1056v1}}].

\bibitem{kravitz}
S.~Kravitz, R.~J. Smith, L.~Hagaman et~al., \emph{Measurements of
  angle-resolved reflectivity of {PTFE} in liquid xenon with {IBEX}},
  \href{https://doi.org/10.1140/epjc/s10052-020-7800-6}{\emph{Eur. Phys. J. C}
  {\bfseries 80} (2020) 262}
  [\href{https://arxiv.org/abs/1909.08730v4}{{\ttfamily 1909.08730v4}}].

\bibitem{jortner}
J.~Jortner, L.~Meyer, S.~A. Rice et~al., \emph{Localized excitations in
  condensed {Ne}, {Ar}, {Kr}, and {Xe}},
  \href{https://doi.org/10.1063/1.1695927}{\emph{The Journal of chemical
  physics} {\bfseries 42} (1965) 4250}.

\bibitem{meg}
{\scshape MEG} collaboration, A.~M. Baldini et~al., \emph{Search for the lepton
  flavour violating decay $\mu ^+ \rightarrow \mathrm {e}^+ \gamma $ with the
  full dataset of the {MEG} experiment},
  \href{https://doi.org/10.1140/epjc/s10052-016-4271-x}{\emph{The European
  Physical Journal C} {\bfseries 76} (2016) 434}
  [\href{https://arxiv.org/abs/1605.05081v3}{{\ttfamily 1605.05081v3}}].

\bibitem{exo}
{\scshape EXO-200} collaboration, G.~Anton et~al., \emph{Search for
  neutrinoless double- $\upbeta$ decay with the complete {EXO}-200 dataset},
  \href{https://doi.org/10.1103/physrevlett.123.161802}{\emph{Physical Review
  Letters} {\bfseries 123} (2019) }
  [\href{https://arxiv.org/abs/1906.02723v3}{{\ttfamily 1906.02723v3}}].

\bibitem{pandaxIII}
{\scshape PandaX-III} collaboration, X.~Chen et~al., \emph{{PandaX}-{III}:
  Searching for neutrinoless double beta decay with high pressure $^{136}${Xe}
  gas time projection chambers},
  \href{https://doi.org/10.1007/s11433-017-9028-0}{\emph{Science China Physics,
  Mechanics {\&} Astronomy} {\bfseries 60} (2017) }
  [\href{https://arxiv.org/abs/1610.08883v2}{{\ttfamily 1610.08883v2}}].

\bibitem{xe1t}
{\scshape XENON} collaboration, E.~Aprile et~al., \emph{Dark matter search
  results from a one ton-year exposure of {XENON1T}},
  \href{https://doi.org/10.1103/physrevlett.121.111302}{\emph{Physical Review
  Letters} {\bfseries 121} (2018) }
  [\href{https://arxiv.org/abs/1805.12562v2}{{\ttfamily 1805.12562v2}}].

\bibitem{lux}
{\scshape LUX} collaboration, D.~Akerib et~al., \emph{Improved limits on
  scattering of weakly interacting massive particles from reanalysis of 2013
  {LUX} data},
  \href{https://doi.org/10.1103/physrevlett.116.161301}{\emph{Physical Review
  Letters} {\bfseries 116} (2016) }
  [\href{https://arxiv.org/abs/1512.03506v3}{{\ttfamily 1512.03506v3}}].

\bibitem{pandax}
{\scshape PandaX-II} collaboration, X.~Cui et~al., \emph{Dark matter results
  from 54-ton-day exposure of {PandaX}-{II} experiment},
  \href{https://doi.org/10.1103/physrevlett.119.181302}{\emph{Physical Review
  Letters} {\bfseries 119} (2017) }
  [\href{https://arxiv.org/abs/1708.06917v2}{{\ttfamily 1708.06917v2}}].

\bibitem{tpc}
M.~Schumann, \emph{Dual-phase liquid xenon detectors for dark matter searches},
  \href{https://doi.org/10.1088/1748-0221/9/08/c08004}{\emph{Journal of
  Instrumentation} {\bfseries 9} (2014) C08004}.

\bibitem{discr}
E.~Aprile, C.~E. Dahl, L.~de~Viveiros et~al., \emph{Simultaneous measurement of
  ionization and scintillation from nuclear recoils in liquid xenon for a dark
  matter experiment},
  \href{https://doi.org/10.1103/physrevlett.97.081302}{\emph{Physical Review
  Letters} {\bfseries 97} (2006) }
  [\href{https://arxiv.org/abs/astro-ph/0601552v1}{{\ttfamily
  astro-ph/0601552v1}}].

\bibitem{xe1tlowmass}
{\scshape XENON} collaboration, E.~Aprile et~al., \emph{Light dark matter
  search with ionization signals in {XENON1T}},
  \href{https://doi.org/10.1103/physrevlett.123.251801}{\emph{Physical Review
  Letters} {\bfseries 123} (2019) }
  [\href{https://arxiv.org/abs/1907.11485v2}{{\ttfamily 1907.11485v2}}].

\bibitem{Aprile:2019jmx}
{\scshape XENON} collaboration, E.~Aprile et~al., \emph{Search for light dark
  matter interactions enhanced by the migdal effect or bremsstrahlung in
  {XENON1T}},
  \href{https://doi.org/10.1103/physrevlett.123.241803}{\emph{Physical Review
  Letters} {\bfseries 123} (2019) }
  [\href{https://arxiv.org/abs/1907.12771v3}{{\ttfamily 1907.12771v3}}].

\bibitem{goodmanwitten}
M.~W. Goodman and E.~Witten, \emph{Detectability of certain dark-matter
  candidates}, \href{https://doi.org/10.1103/physrevd.31.3059}{\emph{Physical
  Review D} {\bfseries 31} (1985) 3059}.

\bibitem{review}
M.~Schumann, \emph{Direct detection of {WIMP} dark matter: concepts and
  status}, \href{https://doi.org/10.1088/1361-6471/ab2ea5}{\emph{Journal of
  Physics G: Nuclear and Particle Physics} {\bfseries 46} (2019) 103003}
  [\href{https://arxiv.org/abs/1903.03026v2}{{\ttfamily 1903.03026v2}}].

\bibitem{Fyield}
G.~N. Vlaskin, Y.~S. Khomyakov and V.~I. Bulanenko, \emph{Neutron yield of the
  reaction ($\upalpha$, n) on thick targets comprised of light elements},
  \href{https://doi.org/10.1007/s10512-015-9933-5}{\emph{Atomic Energy}
  {\bfseries 117} (2015) 357}.

\bibitem{heusser}
G.~Heusser, \emph{Low-radioactivity background techniques},
  \href{https://doi.org/10.1146/annurev.ns.45.120195.002551}{\emph{Annual
  Review of Nuclear and Particle Science} {\bfseries 45} (1995) 543}.

\bibitem{xent}
{\scshape XENON} collaboration, E.~Aprile et~al., \emph{Physics reach of the
  {XENON1T} dark matter experiment},
  \href{https://doi.org/10.1088/1475-7516/2016/04/027}{\emph{Journal of
  Cosmology and Astroparticle Physics} {\bfseries 2016} (2016) 027}
  [\href{https://arxiv.org/abs/1512.07501v2}{{\ttfamily 1512.07501v2}}].

\bibitem{xe100instr}
{\scshape XENON} collaboration, E.~Aprile et~al., \emph{The {XENON}100 dark
  matter experiment},
  \href{https://doi.org/10.1016/j.astropartphys.2012.01.003}{\emph{Astroparticle
  Physics} {\bfseries 35} (2012) 573}
  [\href{https://arxiv.org/abs/1107.2155v2}{{\ttfamily 1107.2155v2}}].

\bibitem{pandaxinstr}
{\scshape PandaX} collaboration, X.~Cao et~al., \emph{{PandaX}: a liquid xenon
  dark matter experiment at {CJPL}},
  \href{https://doi.org/10.1007/s11433-014-5521-2}{\emph{Science China Physics,
  Mechanics {\&} Astronomy} {\bfseries 57} (2014) 1476}
  [\href{https://arxiv.org/abs/1405.2882v1}{{\ttfamily 1405.2882v1}}].

\bibitem{xe1tinstr}
{\scshape XENON} collaboration, E.~Aprile et~al., \emph{The {XENON1T} dark
  matter experiment},
  \href{https://doi.org/10.1140/epjc/s10052-017-5326-3}{\emph{The European
  Physical Journal C} {\bfseries 77} (2017) }
  [\href{https://arxiv.org/abs/1708.07051v1}{{\ttfamily 1708.07051v1}}].

\bibitem{darwinwimp}
M.~Schumann, L.~Baudis, L.~Bütikofer et~al., \emph{Dark matter sensitivity of
  multi-ton liquid xenon detectors},
  \href{https://doi.org/10.1088/1475-7516/2015/10/016}{\emph{Journal of
  Cosmology and Astroparticle Physics} {\bfseries 2015} (2015) 016}
  [\href{https://arxiv.org/abs/1506.08309v2}{{\ttfamily 1506.08309v2}}].

\bibitem{darwin}
{\scshape DARWIN} collaboration, J.~Aalbers et~al., \emph{{DARWIN}: towards the
  ultimate dark matter detector},
  \href{https://doi.org/10.1088/1475-7516/2016/11/017}{\emph{Journal of
  Cosmology and Astroparticle Physics} {\bfseries 2016} (2016) 017}
  [\href{https://arxiv.org/abs/1606.07001v1}{{\ttfamily 1606.07001v1}}].

\bibitem{ntmc}
{\scshape XENON} collaboration, \emph{Projected {WIMP} sensitivity of the
  {XENONnT} dark matter experiment}, {\emph{In preparation} (2020) }
  [\href{https://arxiv.org/abs/2007.08796}{{\ttfamily 2007.08796}}].

\bibitem{tsai}
B.~K. Tsai, D.~W. Allen, L.~M. Hanssen et~al., \emph{A comparison of optical
  properties between solid {PTFE} (teflon) and (low density) sintered {PTFE}},
  in \emph{Reflection, Scattering, and Diffraction from Surfaces}, {SPIE},
  2008, \href{https://doi.org/10.1117/12.798138}{DOI}.

\bibitem{WAGENPFEIL2019577}
M.~Wagenpfeil, T.~Ziegler, A.~Fieguth et~al., \emph{Reflectance of
  {VUV}-sensitive {SiPM} surfaces in liquid xenon},
  \href{https://doi.org/https://doi.org/10.1016/j.nima.2018.09.142}{\emph{Nuclear
  Instruments and Methods in Physics Research Section A} {\bfseries 936} (2019)
  577 }.

\bibitem{PMTtests}
E.~Aprile, M.~Beck, K.~Bokeloh et~al., \emph{Measurement of the quantum
  efficiency of hamamatsu {R8520} photomultipliers at liquid xenon
  temperature},
  \href{https://doi.org/10.1088/1748-0221/7/10/p10005}{\emph{Journal of
  Instrumentation} {\bfseries 7} (2012) P10005}
  [\href{https://arxiv.org/abs/1207.5432v1}{{\ttfamily 1207.5432v1}}].

\bibitem{McPDeuterium}
McPherson, ``Manual for model 632 deuterium light source.''

\bibitem{polarization}
O.~Poulsen, \emph{Polarization calibration of a grating spectrometer},
  \href{https://doi.org/10.1364/AO.11.001876}{\emph{Appl. Opt.} {\bfseries 11}
  (1972) 1876}.

\bibitem{window}
D.~F. Heath and P.~A. Sacher, \emph{Effects of a simulated high-energy space
  environment on the ultraviolet transmittance of optical materials between
  1050 {{\AA}} and 3000 {{\AA}}},
  \href{https://doi.org/10.1364/ao.5.000937}{\emph{Appl. Opt.} {\bfseries 5}
  (1966) 937}.

\bibitem{FSprenger}
F.~Sprenger, \emph{Set-up and test of the slow control and calibration of the
  photomultiplier of an apparatus to measure the reflectivity of teflon in
  {LXe} at {VUV} wavelengths}, {Bachelor}'s thesis, WWU M\"unster, 2013.
\newblock
  \href{http://www.uni-muenster.de/Physik.KP/AGWeinheimer/Files/theses/Bachelor_Florian_Sprenger.pdf}{http://www.uni-muenster.de/Physik.KP/AGWeinheimer/Files/theses/Bachelor\_Florian\_Sprenger.pdf}.

\bibitem{Kubelka1931}
P.~Kubelka and F.~Munk, \emph{{Ein Beitrag zur Optik der Farbanstriche}},
  {\emph{Zeitschrift für technische Physik} {\bfseries 12} (1931) 593–601}.

\bibitem{Kubelka1948}
P.~Kubelka, \emph{New contributions to the optics of intensely light-scattering
  materials. part i}, \href{https://doi.org/10.1364/JOSA.38.000448}{\emph{J.
  Opt. Soc. Am.} {\bfseries 38} (1948) 448}.

\bibitem{Kubelka1954}
P.~Kubelka, \emph{New contributions to the optics of intensely light-scattering
  materials. part ii}, \href{https://doi.org/10.1364/JOSA.44.000330}{\emph{J.
  Opt. Soc. Am.} {\bfseries 44} (1954) 330}.

\bibitem{Boroumand}
F.~Boroumand, J.~E. Moser and H.~van~den Bergh, \emph{Quantitative diffuse
  reflectance and transmittance infrared spectroscopy of nondiluted powders},
  \href{https://doi.org/10.1366/0003702924123502}{\emph{Appl. Spectrosc.}
  {\bfseries 46} (1992) 1874}.

\bibitem{Haefner}
J.~Haefner, A.~Neff, M.~Arthurs et~al., \emph{Reflectance dependence of
  polytetrafluoroethylene on thickness for xenon scintillation light},
  \href{https://doi.org/10.1016/j.nima.2017.01.057}{\emph{Nuclear Instruments
  and Methods in Physics Research Section A} {\bfseries 856} (2017) 86–91}
  [\href{https://arxiv.org/abs/1608.01717}{{\ttfamily 1608.01717}}].

\bibitem{xegasn}
A.~Bideau-Mehu, Y.~Guern, R.~Abjean et~al., \emph{Measurement of refractive
  indices of neon, argon, krypton and xenon in the 253.7{\textendash}140.4 nm
  wavelength range. dispersion relations and estimated oscillator strengths of
  the resonance lines},
  \href{https://doi.org/10.1016/0022-4073(81)90057-1}{\emph{Journal of
  Quantitative Spectroscopy and Radiative Transfer} {\bfseries 25} (1981) 395}.

\bibitem{heidelberg}
D.~Cichon, G.~Eurin, F.~Jörg et~al., \emph{Transmission of xenon scintillation
  light through {PTFE}},  \href{https://arxiv.org/abs/2005.02444v2}{{\ttfamily
  2005.02444v2}}.

\bibitem{chepel}
V.~Chepel and H.~Ara{\'{u}}jo, \emph{Liquid noble gas detectors for low energy
  particle physics},
  \href{https://doi.org/10.1088/1748-0221/8/04/r04001}{\emph{Journal of
  Instrumentation} {\bfseries 8} (2013) R04001}.

\end{thebibliography}\endgroup
\bibliographystyle{JHEP}

\end{document}